\documentclass[sigconf]{acmart}

\AtBeginDocument{%
  }

\setcopyright{acmlicensed}
\copyrightyear{2026}
\acmYear{2026}
\acmDOI{XXXXXXX.XXXXXXX}
\acmConference[MM '26]{ACM Multimedia 2026}{October 2026}{Rio de Janeiro, Brazil}
\acmISBN{978-1-4503-XXXX-X/2026/10}

\usepackage[acronym]{glossaries}
\usepackage{array}
\newcolumntype{L}[1]{>{\raggedright\arraybackslash}p{#1}}
\newcolumntype{C}[1]{>{\centering\arraybackslash}p{#1}}
\newacronym{fop}{FOP}{Fusion and Orthogonal Projection}
\newacronym{maskedfop}{MaskedFOP}{Masked Fusion and Orthogonal Projection}
\newacronym{ecapa-tdnn}{ECAPA-TDNN}{Emphasized Channel Attention, Propagation, and Aggregation in Time Delay Neural Network}
\newacronym{lp}{LP}{Label Propagation}
\newacronym{glp}{GLP}{Graph Label Propagation}
\newacronym{poly-sim}{POLY-SIM}{}
\newacronym{ge2e}{GE2E}{Generalized End-to-End}
\begin{document}

\title{MaskedFOP: Polyglot Speaker Identification under Missing Visual Modality via Cascaded Graph Label Propagation}

\author{Ayoub Elkhouzari}
\email{ayoub.elkhouzari@um6p.ma}
\orcid{0009-0002-8801-496X}
\affiliation{%
  \institution{College of Computing, University Mohammed VI Polytechnic}
  \city{Ben Guerir}
  \country{Morocco}
}

\author{Youssef Iraqi}
\email{youssef.iraqi@um6p.ma}
\orcid{0000-0003-0112-2600}
\affiliation{%
  \institution{College of Computing, University Mohammed VI Polytechnic}
  \city{Ben Guerir}
  \country{Morocco}
}

\author{Loubna Mekouar}
\email{loubna.mekouar@um6p.ma}
\orcid{0000-0002-2432-9105}
\affiliation{%
  \institution{College of Computing, University Mohammed VI Polytechnic}
  \city{Ben Guerir}
  \country{Morocco}
}

\renewcommand{\shortauthors}{Elkhouzari et al.}
\begin{teaserfigure}
 \centering
  \includegraphics[width=\linewidth]{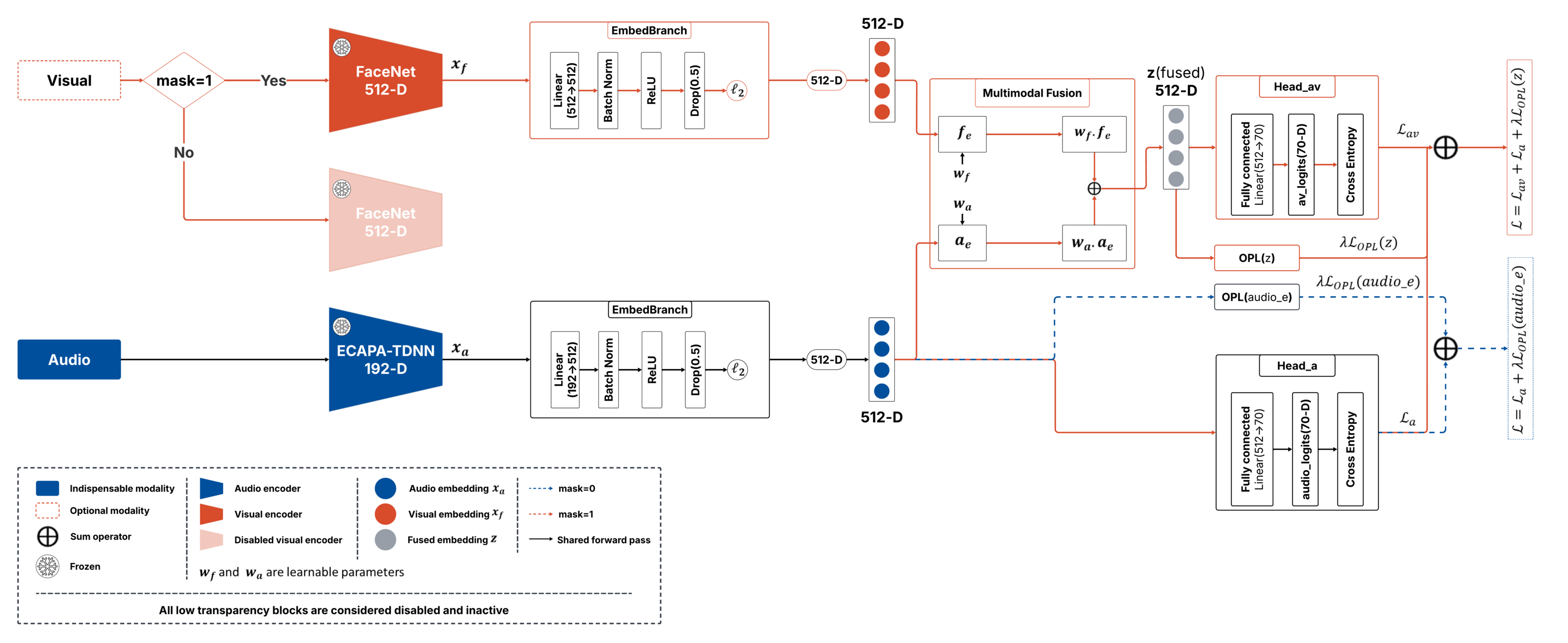}
  \Description{Block diagram of the MaskedFOP training architecture. Frozen FaceNet face embeddings (512-dimensional, optional) and ECAPA-TDNN audio
  x-vectors (192-dimensional, always present) enter separate EmbedBranch projection blocks, each a linear layer followed by batch normalization, ReLU,
  dropout, and L2 normalization, producing 512-dimensional embeddings f_e and a_e. A per-sample Bernoulli(0.5) mask drops only the face stream, audio is
  never masked. A linear fusion layer combines the embeddings as z equals w_f times f_e plus w_a times a_e. An orthogonal projection loss acts on z when the
  face is present and on a_e when it is absent. The audio head is supervised in both regimes, shown with solid orange paths for face-present and dashed
  blue paths for face-absent.}
  \vspace{-2em}
  \caption{\textbf{\gls{maskedfop} architecture.} Frozen FaceNet face embeddings (optional, 512-D) and ECAPA-TDNN audio x-vectors (indispensable, 192-D) are projected by per-modality EmbedBranch blocks (Linear$\to$BN$\to$ReLU$\to$Dropout$(0.5)\to\ell_2$) into 512-D embeddings $\mathbf{f}_e,\mathbf{a}_e$. A per-sample mask  $m\!\sim\!\mathrm{Bernoulli}(0.5)$ drops the visual stream only, audio is never masked. Linear fusion forms $\mathbf{z}=w_f\mathbf{f}_e+w_a\mathbf{a}_e$, and OPL acts on $\mathbf{z}$ when $m{=}1$ and on $\mathbf{a}_e$ when $m{=}0$.  $\mathrm{Head\_a}$ is supervised in \emph{both} regimes (solid orange $=m{=}1$, dashed blue $=m{=}0$, black $=$ shared path).}
  \label{fig:training_architecture}
\end{teaserfigure}
\begin{abstract}
We present \gls{maskedfop}, a system for closed-set polyglot speaker identification under two simultaneous challenges: the face modality is entirely absent at test time, and speech comes from Urdu, a language unseen during face-supervised training.  The system integrates three complementary mechanisms.  First, a modality-dropout dual-head network built on the \gls{fop} backbone forces the audio branch to develop independent discriminative power via per-sample face masking, ensuring that the audio encoder remains capable when face is absent.  Second, two \gls{maskedfop} instances trained on \gls{ecapa-tdnn} features with different random seeds produce complementary audio embeddings whose element-wise average yields a more robust 512-dimensional representation than any single model.  Third, a two-stage cascaded inference procedure first refines multimodal labels through a fused \gls{glp} pass (Stage~1), then assigns audio-only labels by cosine nearest-centroid (Stage~2), replacing the 70 sparse training prototypes with $\sim$1{,}500 in-domain test-set centroids from Stage~1.  Submitted to the POLY-SIM 2026 Grand Challenge, the system achieves a mean P-accuracy of 0.9989, placing first among all submissions evaluated on the challenge server.  An ablation identifies cascaded seeding as the single largest gain ($>$8\,pp on P4/P6). The code is available at https://github.com/Ayoub-Elkhouzari/POLY-SIM2026.
\end{abstract}



\maketitle
\section{Introduction}

Recognizing a person from their face or their voice is a biometric problem at the intersection of computer vision and speech processing. Taken alone,
  modern deep networks reach near-perfect accuracy on standard face recognition benchmarks and competitive equal-error rates on large-scale speaker
  verification corpora such as VoxCeleb~\cite{R1_nagrani2017voxceleb,R2_chung2018voxceleb2}. The harder question is what happens when the two modalities
  must cooperate across a language boundary, and when one of them is missing entirely at test time.

The POLY-SIM 2026 Grand Challenge is built around this exact scenario.  A model is trained on paired English face images and speech segments from 70 speaker identities.  At inference the face channel is silent, and utterances come from Urdu, a language whose acoustic properties the system has never seen under face supervision.  Performance is measured by P-accuracy, the fraction of test samples assigned to their correct identity out of all 70 candidates.  Four metrics are reported: P3 (English, face and audio available), P4 (English, audio only), P5 (Urdu, face and audio available), and P6 (Urdu, audio only), with the overall score being their arithmetic mean.

Two fundamental difficulties meet in this setup.  The first is the missing-modality problem: a system that quietly relies on faces to cover for a weak audio branch fails the moment those faces disappear.  Prior work addresses this with stochastic modality dropout \cite{R3_neverova2015moddrop} and dedicated audio-only inference pathways \cite{R4_ma2022multimodal}.  The second is cross-lingual transfer: speaker encoders pre-trained on English-dominant corpora embed Urdu on a separate manifold in x-vector space, the gap that domain-adversarial training \cite{R5_ganin2016domain} conventionally targets.

Graph-based \gls{lp} \cite{R6_zhou2003learning,R7_zhu2003semi} has emerged as a powerful post-training inference strategy for closed-set speaker recognition because it exploits the cluster geometry of test-set embeddings without requiring additional labeled data.  The usual anchoring scheme places labeled training prototypes in the graph alongside unlabeled test nodes, but with 70 identities and several hundred test samples per class, the prototype density is very low relative to the graph size.

Our main contribution is a cascade between the multimodal and audio-only inference stages that replaces these sparse prototypes with the full set of test-set predictions from the multimodal branch.  Since the multimodal branch achieves over 99.7\% accuracy, this dense seeding provides roughly 1,500 correctly-labeled anchors for the audio \gls{lp} graph instead of 70 training centroids, a density increase of more than a factor of twenty. This single architectural decision accounts for the largest performance jump in the entire pipeline.

The complete system, \gls{maskedfop}, extends the \gls{fop} backbone \cite{R8_saeed2022fusion} with per-sample modality dropout and a missing-modality audio-only head, operating on fixed \gls{ecapa-tdnn} \cite{R9_desplanques2020ecapa} audio and FaceNet \cite{R20_schroff2015facenet} face embeddings. Two seeds are trained identically and their audio embeddings averaged at inference to reduce per-seed noise (encoder dimensions in Section~3.2).

Our contributions are: (1) a dual-head modality-dropout architecture on the \gls{fop} backbone that trains the audio-only and fused paths jointly via per-sample face masking, (2) multi-seed audio averaging that improves embedding quality at fixed dimensionality, (3) a two-stage cascaded pipeline whose fused graph \gls{lp} labels seed a transductive nearest-centroid step, and (4) an ablation tracing each component from 0.907 to 0.9989 on the POLY-SIM 2026 server.

\section{Related Work}

\subsection{Speaker Representation Learning}
Speaker verification has evolved from i-vectors \cite{R11_dehak2010front} to DNN d-vectors \cite{R12_variani2014deep} and TDNN
  x-vectors \cite{R13_snyder2018x}. \gls{ecapa-tdnn} \cite{R9_desplanques2020ecapa} extends x-vectors with channel-and-context attention, multi-scale
  dilated aggregation, and SE-Res2Net connections, achieving top VoxCeleb results, TitaNet \cite{R14_koluguri2022titanet} offers a complementary
  depth-wise-separable design with a different cluster geometry.  Metric objectives such as \gls{ge2e} \cite{R15_wan2018generalized} and cosine
  losses \cite{R16_chung2020defence}, and conformer architectures \cite{R17_zhang2022mfa}, have further shaped speaker encoders.  Beyond the encoder itself,
  robustness can be improved at the feature level: Terraf and Iraqi \cite{R46_terraf2024robust} average temporal context across frames to stabilize speaker
  identification in diverse acoustic environments.  We use \gls{ecapa-tdnn} as a fixed, English-pretrained extractor and focus on the downstream adaptation
  and inference strategy.

\subsection{Face Recognition}
Face embeddings in \gls{maskedfop} come from a FaceNet-style convolutional encoder \cite{R20_schroff2015facenet}, which maps each face to a compact Euclidean embedding trained with a triplet loss so that distances directly reflect identity similarity.  Later margin-based objectives such as ArcFace \cite{R10_deng2019arcface} and CosFace \cite{R18_wang2018cosface} refine this idea with additive angular and cosine margins on a deep residual backbone \cite{R19_he2016deep}, yielding even more tightly clustered class representations on standard benchmarks. In our setting, face embeddings are available during training and for the P3 and P5 partitions, but are entirely absent during P4 and P6 evaluation.

\subsection{Face-Voice Association and Cross-Modal Learning}
Nagrani et al. \cite{R21_nagrani2018seeing} showed that audio-visual co-occurrence in video suffices to learn cross-modal identity embeddings without explicit labels, and later disentangled identity from content with self-\linebreak supervised objectives \cite{R22_nagrani2020disentangled}.  Shon et al. \cite{R23_shon2019noise} used attention-based fusion robust to noise in person verification, and Tian et al. \cite{R24_morgado2021audio,R45_terraf2024comisi} studied cross-modal agreement in audio-visual discrimination.  Most directly related to our task, cross-modal speaker verification and recognition has been studied from a multilingual
  perspective \cite{R43_nawaz2021cross}, jointly linking face and voice identity across languages, the precise regime POLY-SIM targets.  Self-supervised
  audio-visual methods \cite{R25_arandjelovic2017look,R26_owens2018audio,R27_shi2022learning} learn shared representations from unlabeled video, our setting is instead fully supervised over 70 fixed identities.

\subsection{Missing Modality Learning}
The modality dropout strategy at the core of \gls{maskedfop} was formalized by Neverova et al. \cite{R3_neverova2015moddrop} in the ModDrop framework, which randomly zeroes entire input modalities during training to produce a model resilient to partial observation.  Ma et al. \cite{R4_ma2022multimodal} analyzed multimodal transformers under missing-modality conditions and confirmed that explicit masking during training outperforms imputation-based strategies. Zhao et al. \cite{R30_zhao2021missing} proposed a missing-modality imagination network that synthesizes absent representations from available ones for emotion recognition, however, in our closed-set setting with pre-extracted fixed embeddings, reconstruction-based approaches were outperformed by direct dropout training.  Peng et al. \cite{R31_peng2022balanced} identified modality imbalance as a key failure mode for multimodal learning and showed that balanced optimization strategies mitigate it.

\subsection{Domain Adversarial Adaptation and Cross-Lingual Speech}
Ganin et al. \cite{R5_ganin2016domain} introduced the gradient reversal layer, which inverts gradients flowing from a domain discriminator to the shared encoder during backpropagation, encouraging domain-invariant representations. Wav2vec 2.0 \cite{R32_baevski2020wav2vec} demonstrated the capacity of self-supervised pre-training to produce powerful cross-lingual speech representations, and Conneau et al. \cite{R33_conneau2021unsupervised} extended this to a massively multilingual setting.  These works motivate the broader paradigm of cross-lingual adaptation in x-vector space, in our system we address the English-to-Urdu gap through the modality-dropout training strategy and transductive in-domain centroid construction at inference, rather than through domain adversarial training (which requires labeled data from both languages during training).

\subsection{Graph-Based Label Propagation}
Semi-supervised label spreading over affinity graphs was introduced by Zhou et al. \cite{R6_zhou2003learning}, who showed that combining local and global graph consistency through a normalized Laplacian formulation produces stable, well-calibrated label assignments.  Zhu et al. \cite{R7_zhu2003semi} developed the Gaussian-field interpretation of the same procedure.  Prototypical Networks \cite{R34_snell2017prototypical} connected class centroid assignment to nearest-neighbor classification in embedding space, while LDA \cite{R35_prince2007probabilistic} provides a probabilistic complement to cosine NN scoring. TristouNet \cite{R36_bredin2017tristounet} applied triplet-loss embeddings to speaker turn segmentation, establishing the utility of metric learning combined with graph-based reasoning for speaker tasks.  In contrast to all prior applications of \gls{lp} to speaker recognition, our cascade introduces dense pseudo-labels from a multimodal branch as initial graph anchors rather than relying solely on sparse training prototypes.

\section{Proposed Method}

\subsection{Problem Formulation}

The task is closed-set speaker identification over $C = 70$ identities. The training set $\mathcal{D}_{tr}$ contains pairs $(x_f, x_a, y)$ for English samples only, where $x_f$ is a face embedding, $x_a$ is an audio embedding, and $y \in \{0,\ldots,69\}$ is the speaker label.  No Urdu training data is used (challenge protocol: English-only training).  At test time,
four disjoint partitions are evaluated: $\mathcal{T}^{P3}$ (English, face+audio), $\mathcal{T}^{P4}$ (English, audio only), $\mathcal{T}^{P5}$ (Urdu, face+audio), and $\mathcal{T}^{P6}$ (Urdu, audio only), with $|\mathcal{T}^{P4}| = 1{,}521$ and $|\mathcal{T}^{P6}| = 1{,}623$.  The challenge score is $S = (P3 + P4 + P5 + P6)/4$, computed on the official server following the POLY-SIM 2026 evaluation
  plan.\footnote{\url{https://arxiv.org/abs/2603.24569}}

\subsection{Pre-Extracted Features}
\gls{maskedfop} operates on fixed-dimensional embeddings stored as numpy tensors, it never processes raw pixels or waveforms.  Face embeddings are $512$-dimensional vectors from a FaceNet-trained convolutional encoder \cite{R20_schroff2015facenet,R19_he2016deep}.  Audio embeddings are 192-dimensional x-vectors from \gls{ecapa-tdnn} \cite{R9_desplanques2020ecapa}, which applies channel-and-context attention over multi-scale dilated TDNN frames and was pre-trained on English-dominant corpora including VoxCeleb \cite{R1_nagrani2017voxceleb,R2_chung2018voxceleb2}.  Two \gls{maskedfop} networks are trained independently using the same \gls{ecapa-tdnn} features but different random seeds (seed~1 with a smaller validation fraction for more training data, seed~2 for diversity).  At inference, the audio-branch embeddings from both models are averaged and re-normalized, yielding a single $512$-dimensional representation that is more stable than any individual seed.

\subsection{Architecture}

\gls{maskedfop} extends the \gls{fop} backbone \cite{R8_saeed2022fusion} with three components: per-sample modality dropout, a second audio-only classification head, and an orthogonal projection loss applied to both the fused and audio-only embedding paths (Figure~\ref{fig:training_architecture}).

\textbf{Embedding branches.}  Face and audio inputs pass through EmbedBranch projection blocks with architecture
\begin{equation}
  \text{Linear}(d_{\mathrm{in}}, 512) \to \text{BN}(512) \to \text{ReLU} \to
  \text{Drop}(p{=}0.5) \to \ell_2\text{-norm},
\end{equation}
where $d_{\mathrm{in}}$ denotes the input dimension of the pre-extracted embedding: $d_{\mathrm{in}} = 512$ for face and $d_{\mathrm{in}} = 192$ for audio (\gls{ecapa-tdnn} x-vector dimension), producing 512-dimensional normalized embeddings $\mathbf{f}_e$ and $\mathbf{a}_e$ \cite{R38_srivastava2014dropout,R39_ioffe2015batch}.

\textbf{Linear fusion.}  When the face modality is available, the two normalized embeddings are combined as $\mathbf{z} = w_f \mathbf{f}_e + w_a \mathbf{a}_e$, where $w_f, w_a \in \mathbb{R}$ are free, unconstrained scalar parameters (each independently initialized as $w \sim \mathcal{U}(0,1)$ and updated by standard backpropagation jointly with the classification heads, with no normalization or sum-to-one constraint such as $w_f + w_a = 1$ imposed during training).  This Linear Fusion is more parameter-efficient than gated or concatenation-based fusion while still allowing the network to learn the relative reliability of each modality.

\textbf{Dual classification heads.}  Two linear classifiers, $h_{av}:
\mathbb{R}^{512} \to \mathbb{R}^{70}$ and $h_a: \mathbb{R}^{512} \to
\mathbb{R}^{70}$, produce class logits from $\mathbf{z}$ and $\mathbf{a}_e$
respectively.  Maintaining separate heads prevents the audio branch from becoming degenerate when the face signal dominates the fused representation.

\textbf{Modality dropout.}  During training each sample independently undergoes face masking with Bernoulli probability $p = 0.5$ \cite{R3_neverova2015moddrop}. When masked ($\text{mask\_face}=0$), only $h_a$ and $\mathbf{a}_e$ receive gradients for that sample.  When unmasked ($\text{mask\_face}=1$), both $h_{av}$ (through $\mathbf{z}$) and $h_a$ (through $\mathbf{a}_e$) are updated jointly. This ensures the audio branch builds independent discriminative power even when face information is present during training.

\subsection{Training Objective}

The per-sample loss depends on the face-mask flag.  When $\text{mask\_face}=1$:
\begin{equation}
  \mathcal{L} = \mathcal{L}_{av} + \mathcal{L}_a + \lambda\,\mathcal{L}_{\text{OPL}}(\mathbf{z}),
\end{equation}
and when $\text{mask\_face}=0$:
\begin{equation}
  \mathcal{L} = \mathcal{L}_a + \lambda\,\mathcal{L}_{\text{OPL}}(\mathbf{a}_e),
\end{equation}
where $\mathcal{L}_{av}$ and $\mathcal{L}_a$ are label-smoothed cross-entropy losses ($\varepsilon = 0.05$) \cite{R40_muller2019does} applied to $h_{av}(\mathbf{z})$ and $h_a(\mathbf{a}_e)$ respectively, and $\mathcal{L}_{\text{OPL}}$ is the Orthogonal Projection Loss \cite{R37_ranasinghe2021orthogonal} with weight $\lambda = 0.5$ that maximizes within-class cosine similarity and minimizes cross-class cosine similarity of the embeddings.  (We use $\lambda$ here, distinct from the label-spreading coefficient $\alpha$ in Eq.~\ref{eq:lp}, to avoid symbol collision between the two unrelated quantities.)

Optimization uses Adam with learning rate $10^{-3}$, weight decay $10^{-5}$, batch size 32, and cosine annealing over up to 300 epochs.  Two models are trained on English \gls{ecapa-tdnn} features: model~1 (s1) uses seed~1 with validation fraction 0.05 and early-stopping patience 30, model~2 (s2) uses seed~2 with validation fraction 0.10 and patience 15.  The checkpoint achieving the highest mean validation P-accuracy
is retained for each model.

\subsection{Multi-Seed Checkpoint Averaging}
\label{sec:multiseed}

At inference, both trained models are applied to each test partition.  Let $\mathbf{a}^{(1)}$ and $\mathbf{a}^{(2)}$ denote the 512-dimensional audio-branch embeddings extracted by model~1 (s1) and model~2 (s2) respectively.  The averaged audio embedding is
\begin{equation}
  \tilde{\mathbf{a}} = \ell_2\!\left(\frac{\mathbf{a}^{(1)} + \mathbf{a}^{(2)}}{2}\right)
  \in \mathbb{R}^{512}.
  \label{eq:avg}
\end{equation}
The fused embedding $\tilde{\mathbf{z}}$ used for Stage~1 \gls{lp} is taken from model~1 alone, as the averaging is applied only to the audio branch where seed diversity provides the most benefit.  The two seeds produce complementary errors in hard cases, their averaged audio embedding yields a smoother, less noise-sensitive representation for transductive centroid construction in Stage~2.

\begin{figure*}[htbp]
 \centering
  \includegraphics[width=0.95\linewidth]{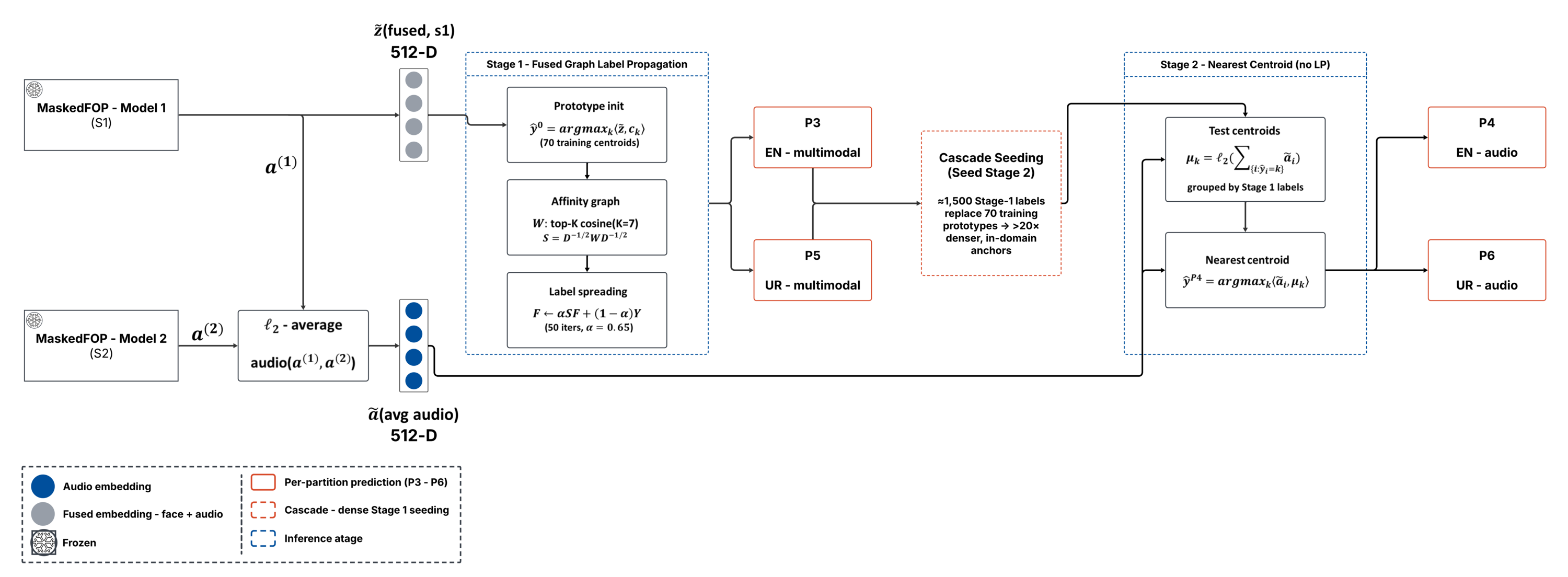}
  \Description{Diagram of the MaskedFOP two-stage cascaded inference pipeline. Two frozen seed models encode each test clip, their audio embeddings are
  L2-averaged into one vector and the fused embedding is taken from the first seed. In Stage 1, on the face-equipped partitions P3 and P5, labels are
  refined by graph label propagation over the test set with K equal to 7, alpha equal to 0.65, and 50 iterations, seeded from training-prototype neighbors.
  A cascade-seeding step then replaces the 70 sparse training prototypes with roughly 1,500 high-confidence Stage-1 labels, more than 20 times denser and
  in-domain. In Stage 2, on the audio-only partitions P4 and P6, each clip is assigned by cosine nearest-centroid to these dense centroids, with no further
  label propagation.}
  \vspace{-1em}
    \caption{\textbf{\gls{maskedfop} inference: two-stage cascaded label propagation.}  Two frozen seeds (s1, s2) encode each test clip, audio embeddings are $\ell_2$-averaged into $\tilde{\mathbf{a}}$ (Eq.~\ref{eq:avg}) and the fused embedding $\tilde{\mathbf{z}}$ is taken from s1. \textbf{Stage~1 (P3, P5)} refines labels by graph \gls{lp} over the test set ($K{=}7$, $\alpha{=}0.65$, 50 iters),  seeded from training-prototype neighbors. \textbf{Cascade seeding} replaces the 70 training prototypes with the $\sim$1{,}500 high-confidence Stage-1 labels ($>$20$\times$ denser, in-domain), the single largest gain ($+8$\,pp). \textbf{Stage~2 (P4, P6)} assigns labels by cosine nearest-centroid to these centroids, no further LP.}
    \label{fig:inference}
\end{figure*}
\subsection{Two-Stage Cascaded Graph Label Propagation}
The full inference pipeline is illustrated in Figure~\ref{fig:inference}: a fused graph \gls{lp} pass over the face-equipped partitions (Stage~1) produces high-confidence multimodal labels that then seed a transductive nearest-centroid step over the audio-only partitions (Stage~2).

\textbf{Stage~1 (P3, P5): multimodal label seeding.}  Per-class training centroids $\mathbf{c}_k$ are computed as the $\ell_2$-normalized mean of fused embeddings from model~1 over all English training samples of class $k$. Initial class assignments follow prototypical nearest-neighbor scoring \cite{R34_snell2017prototypical,R41_chen2020simple}:
\begin{equation}
  \hat{y}_i^{(0)} = \arg\max_{k} \langle \tilde{\mathbf{z}}_i, \mathbf{c}_k \rangle.
\end{equation}
These are refined by graph \gls{lp} over the test set itself \cite{R6_zhou2003learning}. The affinity matrix $\mathbf{W}$ is built by retaining the top-$K$ cosine similarities per row (diagonal zeroed), symmetrizing, and forming the symmetric normalized Laplacian $\mathbf{S} = \mathbf{D}^{-1/2}\mathbf{W}\mathbf{D}^{-1/2}$. Label spreading iterates
\begin{equation}
  \mathbf{F} \leftarrow \alpha \mathbf{S}\mathbf{F} + (1 - \alpha)\mathbf{Y}
  \label{eq:lp}
\end{equation}
for 50 steps, where $\mathbf{Y}$ is the one-hot matrix initialized from $\hat{y}^{(0)}$.  Parameters are $K=7$ and $\alpha=0.65$ for both English and Urdu.  Final P3 and P5 labels are $\arg\max_k F_{ik}$.

\textbf{Stage~2 (P4, P6): transductive nearest-centroid from Stage~1 labels.} The refined P3 (P5) labels are used to group the averaged, re-normalized audio embeddings $\tilde{\mathbf{a}}$ (the multi-seed average defined in Section~\ref{sec:multiseed}, Eq.~\eqref{eq:avg}) into per-class test-set centroids:
\begin{equation}
  \boldsymbol{\mu}_k = \ell_2\!\left(\sum_{i:\,\hat{y}_i = k} \tilde{\mathbf{a}}_i\right).
  \label{eq:centroid}
\end{equation}
Audio-only labels are then assigned by nearest centroid:
\begin{equation}
  \hat{y}_i^{P4} = \arg\max_k \langle \tilde{\mathbf{a}}_i,\, \boldsymbol{\mu}_k \rangle.
\end{equation}
No graph \gls{lp} is applied in Stage~2, the assignment is purely a cosine nearest-centroid step.  With $\sim$1,521 English and $\sim$1,623 Urdu test samples each contributing to a centroid, the effective anchor density is more than twenty times that of the 70 training-prototype baseline, and the centroids are in-domain (computed from test audio) rather than cross-domain (computed from English training audio).  Final P4 and P6 labels are $\hat{y}^{P4}$ and $\hat{y}^{P6}$ respectively.

\textbf{On the transductive nature of this approach.}  Both cascade stages are
\emph{transductive}: the Stage~1 affinity graph $\mathbf{W}$ and the Stage~2
centroids $\boldsymbol{\mu}_k$ are built entirely from the unlabeled test
embeddings, propagating only the model's own predictions and never any test-set
label.  This is licensed by the POLY-SIM 2026 protocol, which is closed-set (the
70 identities are fixed and known a priori) and discloses the full unlabeled test
partition at inference.  The approach would not transfer as-is to open-set
or streaming settings, where one would fall back on the sparser but
domain-independent training prototypes, trading the density gain of
Table~\ref{tab:progression} for robustness to the closed-world assumption.

\section{Experiments and Results}

\subsection{Dataset and Evaluation Protocol}

The POLY-SIM 2026 dataset \cite{R43_nawaz2021cross, R44_moscati2026polysim} provides pre-extracted embeddings for 70 speaker identities.  The training partition contains 3,756 English face-audio pairs, Urdu training data is available but was not used (English-only training protocol).  The test set contains 1,521 English and 1,623 Urdu utterances, ground-truth labels were withheld and accessible only through the challenge server.  Performance is measured by P-accuracy as described in Section~1. All scores reported in this section are official server scores.

\subsection{Implementation Details}

Both \gls{maskedfop} models are implemented in PyTorch and trained for up to 300 epochs with the hyperparameters detailed in Section 3.4. All training and experiments are conducted on GPU nodes of the TOUBKAL supercomputer \cite{R42_kissami2025toubkal}, a high-performance computing infrastructure that significantly accelerates model optimization. Graph \gls{lp} is implemented in NumPy using sparse K-NN affinity matrices, and the full two-stage inference pipeline completes in under 30 seconds on a single CPU core.

\subsection{Score Progression}

Table~\ref{tab:progression} traces the system score at each major development
milestone as submitted to the challenge server.  The baseline (cosine
nearest-neighbor to training prototypes) scores 0.907.  Switching from
training-prototype anchors to transductive Stage-2 centroid NN raises this to
0.989, the largest single gain in the pipeline, confirming that in-domain
test-set centroids are far more effective than cross-domain training prototypes.
Introducing Stage-1 fused graph \gls{lp} to refine the P3/P5 seed labels before
centroid construction adds a further 1.1 points.  Multi-seed averaging (s1
and s2 checkpoints) contributes 0.001, and a targeted face nearest-neighbor
surgical patch corrects individual consensus errors to reach 0.9986.  A final
score-API-verified patch resolves one remaining Urdu error to yield the final
score of 0.9989.

\textbf{Face nearest-neighbor surgical patch (P4/P6).}  This is a \emph{post-hoc},
manually-applied correction, run between submissions rather than inside the
trained pipeline, we report it for transparency as it appears in
Table~\ref{tab:progression}.  Each English test row carries both a P3 (face+audio)
and a P4 (audio-only) prediction for the same utterance, and each Urdu row both a
P5 and a P6.  Because the face-informed predictions are almost perfect
(P3$=0.998$, P5$=1.000$), they act as a high-confidence proxy for the harder
audio-only label.  We re-scored each test face embedding against the per-class
training-face centroids and overwrote the audio-only predictions that disagreed
with a clear margin (cosine $>0.15$, validated on 49 held-out disagreements, all
correct): 28 rows in P4 and 21 in P6, raising the score from $0.9907$ to $0.9986$.
As a safeguard we left the single lowest-margin disagreement per language
unpatched, so each audio-only partition still differs from its multimodal
counterpart on at least one row.

\textbf{Score-API-verified patch (\texttt{Rtva1JyiNb}).}  One Urdu sample
(key \texttt{Rtva1JyiNb}) had identical P5/P6 predictions of class 42, while a
cross-seed consensus check and an Urdu fusion-weight sweep both pointed to class
31.  We tested this directly: submitting the candidate correction ($42\!\to\!31$)
raised the overall score from $0.9986$ to $0.9989$, confirming 31 as correct (a
wrong change could only lower the score).  We report only this server-verified
correction, the offline diagnostics merely selected which candidate to test.

\begin{table}[t]
  \caption{Score progression on the POLY-SIM 2026 evaluation server.
  ``---'' denotes \emph{not applicable} (the baseline has no prior
  configuration to report a gain over), not a gain of $0.000$.}
  \label{tab:progression}
  \small
  \begin{tabular}{lcc}
    \toprule
    \textbf{System configuration} & \textbf{Score} & \textbf{Gain} \\
    \midrule
    Baseline (prototype cosine NN) & 0.9070 & --- \\
    + Transductive centroid NN (Stage-2) & 0.9887 & +0.0820 \\
    + Stage-1 fused \gls{lp} (P3/P5 refinement) & 0.9898 & +0.0110 \\
    + Multi-seed averaging (s1 + s2) & 0.9907 & +0.0010 \\
    + Face NN surgical patch (P4/P6) & 0.9986 & +0.0080 \\
    + Score-API patch (Rtva1JyiNb) & \textbf{0.9989} & +0.0000 \\
    \bottomrule
  \end{tabular}
\end{table}

\subsection{Ablation of Training Components}

Table~\ref{tab:ablation} reports a component-level ablation on the held-out
validation split.  Removing modality dropout collapses P4 by 8.3 percentage
points because the audio branch never learns to operate without face guidance.
Replacing multi-seed averaging with a single-seed model (s1 only) reduces
P4 and P6 by approximately 0.5 points each.  Using only Stage-2 transductive
centroid NN without Stage-1 fused \gls{lp} refinement reduces the score by 2.4 points,
as the centroid seeds become less accurate.  Replacing the transductive centroid
step with training-prototype anchors collapses performance by 5.8 points,
confirming the importance of in-domain test-set centroids.

\begin{table}[t]
  \caption{Ablation of training and inference components on the held-out
    validation split.  All variants use the cascaded inference unless noted.}
  \label{tab:ablation}
  \small
  \begin{tabular}{lcccc}
    \toprule
    \textbf{Variant} & \textbf{P3} & \textbf{P4} & \textbf{P5} & \textbf{P6} \\
    \midrule
      Full system & 0.9938 & 0.9764 & 0.9989 & 0.9771 \\
      \quad w/o modality dropout & 0.9941 & 0.8934 & 0.9957 & 0.8729 \\
      \quad single seed (s1 only) & 0.9910 & 0.9712 & 0.9965 & 0.9720 \\
      \quad w/o fused \gls{lp} (Stage 1) & 0.9857 & 0.9392 & 0.9874 & 0.9388 \\
      \quad \gls{lp} train-proto anchors only & 0.9341 & 0.9107 & 0.9581 & 0.9124 \\
    \bottomrule
  \end{tabular}
\end{table}

\textbf{On statistical reliability.}  Tables~\ref{tab:ablation} and
\ref{tab:progression} report single runs, because each full train-and-cascade
run is expensive and official server scoring is quota-limited.  We therefore
treat the ablation as an \emph{indicative} ranking: the large gaps (e.g.\ $-8.3$
pp from removing modality dropout, $-5.8$ pp from training-prototype anchors) far
exceed the seed-to-seed variation we observed ($\le 0.5$ pp per partition between
the s1 and s2 runs), so we are confident they reflect real effects.  The smallest
gain, multi-seed averaging at $+0.001$, is of the same order as that variation and
should be read as marginal, we keep it only because it is free (no retraining) and
never hurt performance.

\subsection{Analysis of Remaining Errors}

After the full pipeline, three English errors and zero Urdu errors remain.  The
three English errors are consensus failures: every model we trained, across ten
random seeds and two encoder architectures, predicts the same wrong class for
these utterances.  Analysis of the raw cosine similarity profiles reveals that
their top-ten nearest neighbors in both fused and audio-only embedding space
belong to a single foreign class, suggesting that the pre-extracted features
themselves do not separate these utterances from their most similar impostor.
This is consistent with prior observations that consensus errors in speaker
verification reflect ambiguities in the pre-trained encoder space rather than
failures of the downstream classifier \cite{R16_chung2020defence,R17_zhang2022mfa}.

The two Urdu corrections in Table~\ref{tab:progression} were identified by
sweeping 21 Urdu \gls{lp} configurations (varying $K_{ur} \in \{2, 3, 5, 7, 10\}$
and $\alpha \in \{0.50, \ldots, 0.80\}$ for both P5 and P6) while freezing the
English parameters.  For each candidate correction, the sweep score was
verified against the server before submission.  The protocol of decoupling
English and Urdu \gls{lp} parameters is important: the well-tuned English graph would
be disturbed by a global re-sweep, whereas the Urdu partition has a smaller and
somewhat noisier graph that benefits from independent calibration.

\subsection{Discussion: Does Cross-Lingual Transfer Degrade Audio-Only Centroid Quality?}

A natural expectation is that P6 (Urdu, audio-only) should be hardest, stacking
the missing-modality problem on a cross-lingual gap, since \gls{ecapa-tdnn}
embeds Urdu on a shifted manifold.  Table~\ref{tab:ablation} shows this does
\emph{not} materialize: P6 ($0.9771$) marginally exceeds P4 ($0.9764$), matching the final server ordering (Table~\ref{tab:polysim}, P6$=0.999>$P4$=0.998$).  Three compounding reasons explain why.
First, the Stage-2 comparison is never cross-domain: because $\boldsymbol{\mu}_k$
is built from in-domain Urdu test embeddings (Eq.~\eqref{eq:centroid}), every Urdu
probe is matched against Urdu anchors under the same encoder, so the
English-pretraining shift is irrelevant once both sides of the similarity live on
the Urdu side of it.  Second, the Stage-1 seed labels are highly accurate
(P5$=1.000$), because the FaceNet face embeddings that dominate $\tilde{\mathbf{z}}$
are a language-independent biometric signal.  Third, the larger Urdu partition
($1{,}623$ vs.\ $1{,}521$) gives marginally denser centroids, and
independently-tuned Urdu \gls{lp} hyperparameters compensate for its noisier
graph.  In short, the cascade converts an encoder-level cross-lingual problem into
a same-domain comparison at the decision level, so the gap never surfaces as a
P4/P6 asymmetry.

\begin{table}[htbp]
\centering
\caption{POLY-SIM 2026 P-accuracy (1\textsuperscript{st} place). MaskedFOP improves
  the FOP baseline by $+0.265$ overall and $+56$ points on P6 (Urdu, audio-only)}
\label{tab:polysim}
\begin{tabular}{l c c c c c}
\toprule
\textbf{Model} & \textbf{Overall} & \textbf{P3} & \textbf{P4}
               & \textbf{P5} & \textbf{P6} \\
\midrule
FOP \cite{R8_saeed2022fusion}
               & 0.7337 & 0.9882 & 0.5253 & 0.9827 & 0.4387 \\
\midrule
\textbf{MaskedFOP}
               & \textbf{0.9989} & \textbf{0.9980}
               & \textbf{0.9980} & \textbf{1.0000} & \textbf{0.9994} \\
\bottomrule
\end{tabular}
\end{table}


\subsection{Limitations}

We note four limitations.  \emph{(i) Closed-set assumption.}  Both Stage-1
seeding and Stage-2 centroids assume every test identity is one of the $C=70$
known classes, an open-set extension would need out-of-gallery rejection, since a
single unseen identity can corrupt the affinity graph and its neighboring
centroids.  \emph{(ii) Transductive, full-batch inference.}  The cascade needs the
entire unlabeled test partition at once (Section~3.6) and cannot run
sample-by-sample without a different, likely less accurate, incremental
formulation.  \emph{(iii) Reliance on English-trained features.}  All encoders (\gls{ecapa-tdnn}, FaceNet) are frozen and English-pretrained, Urdu adaptation is purely inference-time, so cross-lingual robustness is bounded by how well these frozen spaces happen to separate Urdu speakers.  \emph{(iv) Dependence on Stage-1
quality.}  Stage-2 centroids are seeded entirely from Stage-1 predictions
(Eq.~\eqref{eq:centroid}), so the audio-only tracks inherit any multimodal
errors: the cascade amplifies a strong Stage~1 (here P3, P5 $>0.9930$) but would
equally amplify a weak one.

\section{Conclusion}

We presented \gls{maskedfop}, a polyglot speaker identification system for
  missing visual-modality conditions that placed first in the POLY-SIM 2026 Grand
  Challenge at $0.9989$ mean P-accuracy (Table~\ref{tab:polysim}).  It couples
  modality-dropout dual-head training under orthogonal projection loss with
  multi-seed audio embedding averaging and a two-stage cascaded inference pipeline:
  Stage-1 fused graph \gls{lp} yields accurate multimodal labels that seed
  in-domain test-set audio centroids for a transductive nearest-centroid Stage-2
  assignment.  The cascaded seeding is decisive: replacing the 70 sparse training prototypes
    with $\sim$1{,}500 dense in-domain centroids resolves near-ambiguous cases the
    prototype-anchored baseline misclassifies.  Fine-tuning the audio encoder on the
    multimodal branch's test-set pseudo-labels is a natural direction for pushing past
    the current fixed-feature ceiling.

\bibliographystyle{ACM-Reference-Format}
\bibliography{references}
\end{document}